\documentclass[aps,prc,showpacs,showkeys,superscriptaddress,nofootinbib,twocolumn,floatfix]{revtex4}
\usepackage{bm}
\usepackage{placeins}
\usepackage[]{graphicx}
\usepackage{amsfonts}
\usepackage{multirow}
\usepackage{graphicx,color,amsmath,amssymb}

\newcommand\blfootnote[1]{%
  \begingroup
  \renewcommand\thefootnote{}\footnote{#1}%
  \addtocounter{footnote}{-1}%
  \endgroup
}

\newcommand{\eg}{\textit{e.g.}}

\newcommand{\ie}{\textit{i.e.}}

\newcommand{\X}{$X(3872)$}
\newcommand{\Gmol}{\Gamma_{\rm mol}}
\newcommand{\Gtet}{\Gamma_{\rm tet}}
\newcommand{\Gmzero}{\Gamma_0^{\rm mol}}
\newcommand{\Gtzero}{\Gamma_0^{\rm tet}}
\newcommand{\pT}{$p_T$}
\newcommand{\Td}{$T_{\rm diss}$}

\begin{document}

\title{X(3872) Transport in Heavy-Ion Collisions}

\author{Biaogang Wu\footnote{bgwu@tamu.edu}}
\address{Cyclotron Institute and Department of Physics and Astronomy, Texas A$\&$M University, College Station, TX 77843-3366, USA}
\author{Xiaojian Du\footnote{xiaojiandu.physics@gmail.com}}
\address{Fakult\"at fur Physik, Universit\"at Bielefeld, D-33615 Bielefeld, Germany}
\author{Matthew Sibila\footnote{m-sibila@onu.edu}}
\address{Department of Physics $\&$ Astronomy, Ohio Northern University, Ada, OH 45810, USA} 
\author{Ralf Rapp\footnote{rapp@comp.tamu.edu}}
\address{Cyclotron Institute and Department of Physics and Astronomy, Texas A$\&$M University, College Station, TX 77843-3366, USA}

\date{\today}

\begin{abstract}
The production of the \X~particle in heavy-ion collisions has been contemplated as an alternative probe of its internal structure. 
To investigate this conjecture, we perform transport calculations of the \X~through the fireball formed in nuclear collisions at the LHC. 
Within a kinetic-rate equation approach as previously used for charmonia, the formation and dissociation of the \X~is controlled by 
two transport parameters, \ie, its inelastic reaction rate and thermal-equilibrium limit in the evolving hot QCD medium. While the 
equilibrium limit is controlled by the charm production cross section in primordial nucleon-nucleon collisions (together with the 
spectra of charm states in the medium), the structure information is encoded in the reaction rate. We study how different 
scenarios for the rate affect the centrality dependence and transverse-momentum (\pT) spectra of the \X. Larger reaction 
rates associated with the loosely bound molecule structure imply that it is formed later in the fireball evolution than the
tetraquark and thus its final yields are generally smaller by around a factor of two, which is qualitatively different from most 
coalescence model calculations to date. The \pT~spectra provide further information as the later decoupling time within the 
molecular scenario leads to harder spectra caused by the blue-shift from the expanding fireball.
\end{abstract}
\maketitle

\section{Introduction}
\label{sec_intro} 
\blfootnote{\\Published in Eur. Phys. J. A 57, 122 (2021)\\Doi: 10.1140/epja/s10050-021-00435-6}Ever since its discovery in electron-positron annihilation experiments~\cite{Choi:2003ue}, the internal structure of the \X~particle has 
remained under debate (cf., \eg, Ref.~\cite{Esposito:2016noz} for a recent review). While its constituent-quark content is generally 
believed to be of $c\bar c q \bar q$ type (with a charm-anticharm and a light quark-antiquark pair), their internal arrangement is more 
controversial. On the one hand, the proximity of the \X~mass to the threshold of a $D$ and $D^*$ meson is suggestive for a weakly 
bound molecular state~\cite{Hanhart:2007yq,Molina:2009ct}; on the other hand, 
its small decay width appears to suggest that its wave function has little overlap with $DD^*$ configurations, thus favoring a bound state 
of a color-antitriplet diquark ($cq$) and an antidiquark ($\bar c \bar q$)~\cite{Maiani:2004vq,Riek:2010fk}. In principle, also a superposition 
of the two configurations is possible. With the advent of first data on the production of the \X~in heavy-ion collisions (HICs) at the 
LHC~\cite{CMS:2019vma}, a new way of addressing this problem has opened up. Since the different hadronic structures are expected to 
affect how the \X~couples to the surrounding QCD medium, its observable yields in HICs have been conjectured to provide novel insights 
into the nature of this particle.

Thus far, the problem of \X~production in HICs has mainly been addressed using instantaneous coalescence models 
(ICMs)~\cite{Cho:2010db,Sun:2017ooe,Fontoura:2019opw,Zhang:2020dwn}, 
by calculating its yield at the hadronization transition with a suitable 
wave function in coordinate space to encode the different structure information. However, in an instantaneous approximation, energy 
conservation can not be guaranteed which can cause problems in recovering the thermal-equilibrium limit~\cite{Ravagli:2007xx}. As a 
result, variations in the \X~yields of up to two orders of magnitude have been predicted, essentially depending on the assumptions about 
its wave function. On the contrary, in the statistical hadronization model (SHM)~\cite{Andronic:2019wva}, which is based on the assumption 
of thermal equilibrium (with a charm-quark fugacity factor to ensure charm-quark conservation), the \X~yields only depend on its mass, \ie, 
they are independent of internal structure that does not affect the mass.

In the present paper we will revisit \X~production in HICs by conducting a calculation of its transport through the fireball formed 
in Pb-Pb collisions at the LHC. Employing the thermal-rate equation framework that we have relied on in the past to interpret and 
predict a wide variety of charmonium and bottomonium observables~\cite{Grandchamp:2003uw,Zhao:2011cv,Du:2017qkv}, the 
time evolution of the \X~abundance is determined by two transport parameters: the equilibrium limit and the inelastic reaction 
rate. The former provides an important benchmark as the long-time limit of the transport equation, while the latter encodes the structure
effects through its coupling to the medium. In this way we can combine structure information from coalescence (and absorption) 
processes in momentum space with the universal equilibrium limit in a controlled fashion.
Due to the weak binding of the \X~(at least in the molecular scenario), we focus on the effects of the interacting hadronic medium 
which makes up nearly half of the fireball lifetime in central Pb-Pb collisions. Since the binding energy (relative to the nearest hadronic 
threshold) is much smaller than the fireball temperature, medium effects in the evolving hadronic phase are expected to play an 
important role; they are neglected in both the SHM and some of the coalescence model applications to calculate \X~observables in HICs.
The effects of hadronic transport on the \X~have been studied in Refs.~\cite{Cho:2013rpa,Abreu:2016qci} for Au-Au collisions at RHIC. 
In Ref.~\cite{Cho:2013rpa} the hadronic dissociation cross sections for a spin-1 \X~were evaluated from $\pi$- and $\rho$-meson 
induced break-up and found to be small; consequently, no significant effect of the hadronic transport was discerned in both scenarios, 
and the final result essentially reflected the production yields at hadronization as taken from Ref.~\cite{Cho:2010db}, with much larger 
yields for the molecule scenario (similar results were obtained for the doubly-charmed tetraquark, 
$T_{cc\bar q \bar q}$~\cite{Hong:2018mpk}). In Ref.~\cite{Abreu:2016qci}, large reaction rates were inferred for the tetraquark 
scenario, leading to a large suppression so that also here the final yields turned out to be much smaller than for the molecular scenario.
Our present approach, aside from focusing on Pb-Pb collisions at LHC, differs from previous works in several aspects: our initial 
conditions vary between zero and the equilibrium limit (motivated by our previous transport results for charmonia), our reaction rates 
are generally larger (as suggested by recent literature), and the (temperature-dependent) equilibrium limit includes a large number of 
charm-hadron states which largely affects the evaluation of the charm-quark fugacity (which figures squared for states containing 
$c\bar c$). In addition, we also provide a centrality dependence and calculate transverse-momentum ($p_T$) spectra as another 
tool to discriminate production times in the evolution. 

The remainder of this paper is organized as follows. In Sec.~\ref{sec_trans} we briefly recapitulate the main features of our 
kinetic-rate equation approach specifically discussing the new components for the \X~calculation (in particular the transport parameters). 
In Sec.~\ref{sec_evo} we present and discuss our results for the time evolution of the \X~equilibrium limit and the solutions of the rate 
equation for its yield in a molecular vs.~a tetraquark scenario. In Sec.~\ref{sec_obs} we calculate \X~observables in terms of its centrality 
dependence and $p_T$ spectra, thereby addressing whether and how these can be used to discriminate different structure scenarios. 
In Sec.~\ref{sec_concl} we summarize and conclude including a brief discussion of our results in light of other works on \X~production 
available in literature.

\section{Transport Approach}
\label{sec_trans} 
Our transport approach starts from a rate equation~\cite{Grandchamp:2003uw,Zhao:2011cv,Du:2017qkv},
\begin{equation}
\frac{dN_X(\tau)}{d\tau}=-\Gamma(T(\tau))\left[N_X(\tau)-N_X^{\rm eq}(T(\tau))\right],
\label{rate-eq}
\end{equation}
for the number of \X~particles, $N_X$, governed by the two transport parameters, the equilibrium limit, 
\begin{equation}
N_X^{\rm eq}(T) =  V_{\rm FB} 3 \gamma_c^2 \int \frac{d^3k}{(2\pi)^3} \exp(-E_k/T)     \  , 
\end{equation}
($E_k=\sqrt{k^2+m_X^2}$) and reaction rate, $\Gamma$ ($V_{\rm FB}$ denotes the time-dependent volume of the expanding fireball). 
The calculation of the equilibrium limit follows the standard 
charm-conservation condition, 
\begin{equation}
\label{Neq}
N_{c\bar c}=\frac{1}{2}\gamma_{c} n_{\rm{op}}V_{\rm{FB}}\frac{I_1(\gamma_{c} n_{\rm{op}}V_{\rm{FB}})}
{I_0(\gamma_{c} n_{\rm {op}}V_{\rm{FB}})} + \gamma_{c}^2 n_{\rm{hid}} V_{\rm{FB}}  \ , 
\end{equation}
where $N_{c\bar c}$ denotes the number of charm--anticharm-quark pairs in the fireball as determined by the elementary charm cross section 
in proton-proton ($pp$) collisions at given center-of-mass energy, and the number of primordial nucleon-nucleon collisions, $N_{\rm coll}$, in a 
heavy-ion collision at given centrality.
As mentioned in the introduction, the key input to compute the charm-quark fugacity factor, $\gamma_c$, is (other than the charm cross section) 
the spectrum of open-charm states included in the summation of the first term on the right-hand-side ($rhs$) of Eq.~(\ref{Neq}) (in practice, the 
contribution of charmonia (second term on the $rhs$) and multiple-charm hadrons is negligible, and thus the inclusion of exotic $X$ states plays no 
role in the determination of $\gamma_c$ either). 
In the QGP these are simply charm quarks, while in hadronic matter one needs to sum over all available charm hadrons, $\alpha=D, D^*, \Lambda_c,...$ 
and their antiparticles, with their respective masses ($m_\alpha$), 
\begin{equation}
 n_{\rm{op}} = \sum\limits_\alpha n_\alpha(T;m_\alpha) \ .
\end{equation} 
In our previous works~\cite{Grandchamp:2003uw,Zhao:2011cv,Du:2015wha}, we included all charm states listed by the particle data group (PDG), 
together with a 5~TeV $pp$ charm cross section of $d\sigma_{c\bar c}/dy$=0.8\,mb at mid-rapidity. More recent developments suggest that 
charm-baryon production is significantly larger than assumed before~\cite{Acharya:2017kfy}, presumably due to ``missing states" not listed by 
the PDG~\cite{He:2019tik}. However, when including the latter in $n_{\rm op}$ (as we do here), one needs to also account for the increased 
cross section due to the extra states (amounting to $d\sigma_{c\bar c}/dy$$\simeq$1.1\,mb), and the net effect on the fugacity essentially 
cancels out (which we have verified explicitly). Thus, our evaluation of the equilibrium limit of the \X~particle in the hadronic phase appears to 
be rather stable.We also note that we include the correlation volume effects~\cite{Grandchamp:2003uw} in the canonical suppression factor, 
$I_1/I_0$, in Eq.\,(\ref{Neq}). For Pb-Pb collisions at LHC energies they are immaterial except for the most peripheral (60-80\%) centrality bin 
that we consider below, for which we have checked that uncertainties in the modeling of the correlation volume expansion can lead up to a 
40\% reduction in the equilibrium limit of the \X.  Our baseline scenarios will neglect any shadowing of the initial charm-production. However, for 
the centrality-dependent nuclear modification factor of the \X~we will illustrate the effect of a charm cross section which is suppressed by up to 
20\% in central collisions.

For the second transport parameter, the reaction rate (or inelastic width), $\Gamma$, we do not perform an independent microscopic calculation 
but rather take guidance from the literature to define typical ranges and temperature dependencies that represent the molecular ($\Gmol$) and 
tetraquark ($\Gtet$) bound-state structures. In Ref.~\cite{Cleven:2019cre} the \X~width in a pion gas has been calculated within the molecule 
scenario through the dressing of the $D$ and $D^*$ constituents and found to be about $\Gmol\simeq$~60\,MeV at $T$$\simeq$150\,MeV, quite 
consistent with twice the collisional width of $D$-mesons in a pion gas~\cite{Fuchs:2004fh,He:2011yi}. However, the number density of a hadron 
resonance gas is substantially larger than that of a pion gas, by about a factor of $\sim$6-7 at $T_0$=180\,MeV relative to a pion gas at $T$=150\,MeV 
(in addition, interactions with excited states (vector mesons etc.) or anti-/baryons are not subject to a Goldstone suppression). Thus, we assume 
a width range of ${\Gmzero}\simeq (400\pm100)$\,MeV for the molecular scenario at our initial temperature $T_0$. For the tetraquark configuration, 
presumably a diquark--antidiquark, rather little information is available; due to the small overlap of its wavefunction with color-neutral states, its 
hadronic in-medium width is expected to be rather small. For example, when employing the $XDD^*$ coupling estimated from the decay branching 
in the vacuum~\cite{Brazzi:2011fq}, the pion-induced absorption rate of the \X~turns out to be a few MeV at $T$=160\,MeV~\cite{Cho:2013rpa}. 
An upper estimate might be derived from geometric-scaling arguments, \ie, assuming $\sigma_{\rm diss} \simeq \pi r^2$ for a particle radius $r$, with 
a phase space suppression when the collision energy with the medium particle approaches the mass threshold of the outgoing 
particles~\cite{Ferreiro:2018wbd}. If one adopts the recently suggested \X~size of 1.3\,fm (which is at the high end for typical tetraquark 
configurations) together with a total hadron density of $\rho_{\rm had}^{\rm tot}\simeq0.8/{\rm fm}^3$ and $v_{\rm rel}\simeq0.6c$, one 
obtains $\Gtzero \simeq 80$\,MeV as a maximal value. As a conservative (upper) range for the total \X~width in the tetraquark configuration we 
therefore use $\Gtzero$=50-80\,MeV at $T_0$=180\,MeV, which is still almost an order of magnitude smaller than $\Gmzero$.

\section{Time Evolution}
\label{sec_evo}
\begin{figure}[!t]
\begin{minipage}[b]{1.0\linewidth}
\centering
\includegraphics[width=1.0\textwidth]{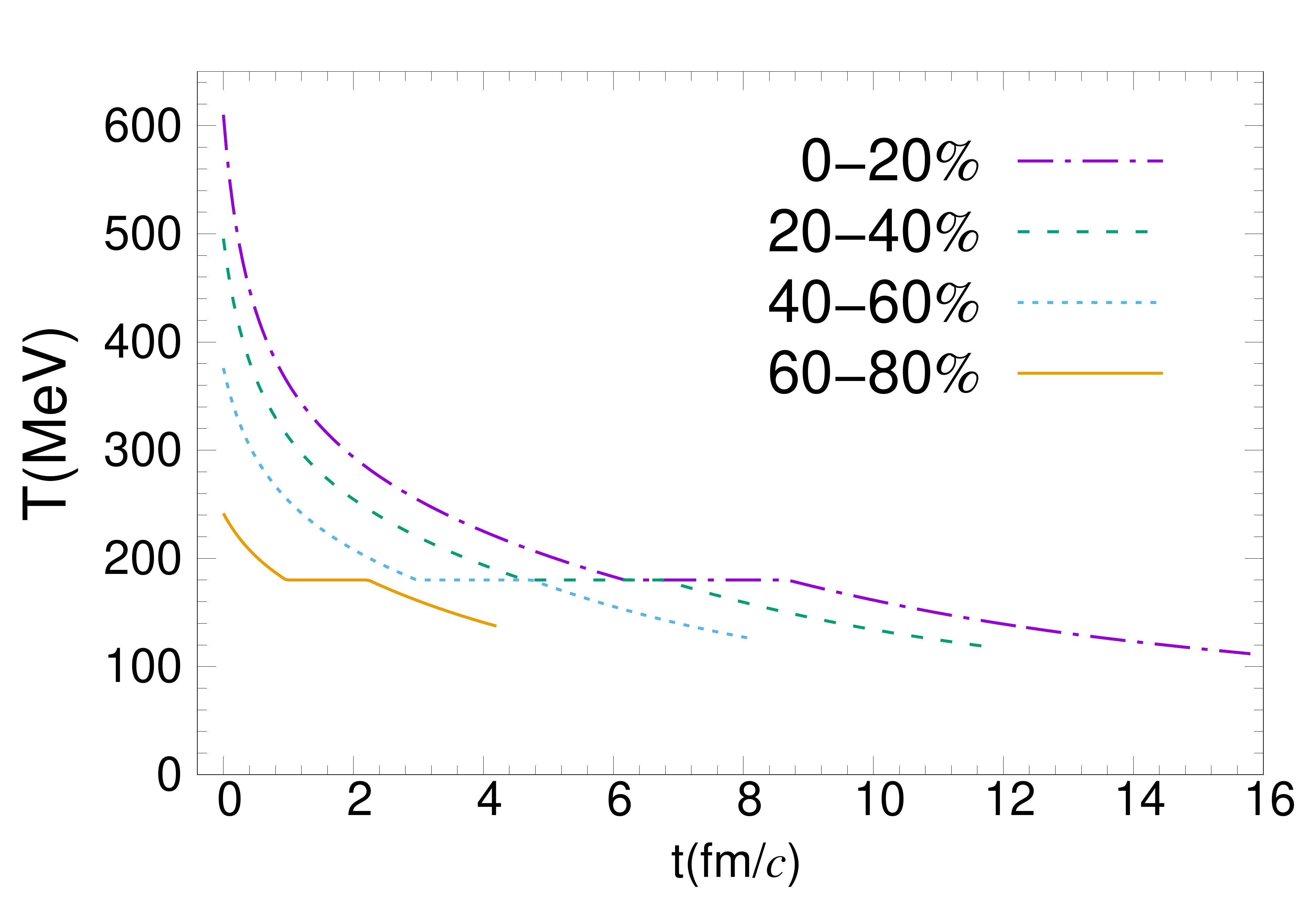}
\end{minipage}
\caption{Time evolution of temperature in the expanding hadronic medium in 5\,TeV Pb-Pb collisions at different centralities within the thermal 
fireball model. Note that for the present investigation, only the hadronic phase evolution figures.} 
\label{fig_temp}
\end{figure}

In this section we inspect the time dependence of various \X-related quantities through the fireball evolution. We will focus on Pb-Pb collisions at 
5\,TeV and employ the same bulk-medium evolution that we have been using for charmonium and bottomonium transport in the 
past~\cite{Grandchamp:2003uw,Zhao:2011cv,Du:2015wha,Du:2017qkv}; it is approximated by a cylindrically expanding fireball volume with a transverse flow 
profile of blastwave type, with evolution parameters that reproduce the fits to empirical light-hadron spectra (pions, kaons protons) at thermal 
freezeout temperatures of  around $T_{\rm fo}$=110\,MeV (somewhat larger for peripheral collisions). We start our hadronic evolution at a 
temperature of $T_0$=180\,MeV (where also chemical freezeout is assumed to occur), but the results will be similar when using an initial 
temperature of $T_0$=170\,MeV with chemical freezeout at $T_{\rm ch}$=160\,MeV, as long as the total entropy at a given centrality is the same, 
as fixed by the observed hadron multiplicities (see also Ref.~\cite{Du:2017qkv} for the case of bottomonium transport). The choice of $T_0$=180\,MeV 
can be considered as an upper limit for the effects of the hadronic evolution. We have also verified for the case at hand that varying the parameters 
in the time evolution of the fire-cylinder, specifically the transverse acceleration of the fireball boundary, by $\pm20$\% has negligible effects on the 
transport yields of the \X.
Similar to the case of the charm-quark fugacity, effective chemical potentials have been introduced after chemical freezeout for hadrons which 
are stable under strong interactions (pion, kaon, nucleons, etc.) to ensure that the experimental observed abundancies are conserved. As mentioned 
above, the key input parameter is the total entropy at a given Pb-Pb collision centrality, which is adjusted to the pertinent light-hadron production yields 
and largely determines the volume and temperature evolution of the fireball. In Fig.~\ref{fig_temp} we 
summarize the time dependence of the fireball's temperature for the 4 centrality bins that we consider in this paper for 5.02\,TeV PbPb collisions.
We note that the use of a quasiparticle QGP equation, matched to a hadron resonance gas at $T_0$=180\,MeV, leads to a plateau in temperature
evolution due a standard mixed phase construction, which, however, does not affect the hadronic evolution used for our \X~transport calculations. 

Next, we introduce the initial conditions for the number of \X~particles at the beginning of the hadronic phase. For the hadronic molecule 
scenario, the weak binding is not expected to produce any bound states prior to the hadronic phase. Our baseline assumption therefore is that 
the initial hadronic abundance of the molecule configuration is zero (we will investigate different ``melting" temperatures in the hadronic phase, 
where regeneration starts). This is quite different from typical ICMs where the assumption of a wave function which is smeared out over several 
femtometers provides a large phase space and thus results in large yields that can markedly exceed the equilibrium 
limit~\cite{Cho:2010db,Zhang:2020dwn} (one may also question whether a large-size hadronic molecule can exist at all in a medium with an 
interparticle spacing that is much smaller than the distance between the molecule's constituents). In our approach, the formation is rather encoded 
in a large but finite reaction rate in the hadronic phase, which, on the contrary, drives the abundance toward the equilibrium limit. For the tetraquark, 
the relatively small reaction rates in the hot and dense hadronic phase suggest that its main production occurs at an earlier stage, \ie, in the 
strongly-coupled  quark-gluon plasma (sQGP) close to 
the ``transition" temperature. In this regime one expects attractive quark-quark (antiquark-antiquark) interactions in the color antitriplet (triplet) 
channel to form strong (anti-) diquark correlations~\cite{Shuryak:2003ja,Riek:2010fk} (as precursors of baryon formation), which in turn can 
further (re-)combine into tetraquarks as diquark--antidiquark bound states. A microscopic transport calculation of these processes is beyond the 
scope of the present paper. Instead, as the reaction rates in the sQGP are generally high, we simply assume that the \X~abundance reaches 
close to its chemical equilibrium value at the beginning of the hadronic phase. This is supported by our previous transport calculations for the 
$\psi'$~\cite{Du:2015wha}, which is a similarly loosely bound state made of colored constituents. The uncertainty in this assumption for the 
initial condition for the \X~should thus be no more than a few 10's of percent.

For the temperature dependence of the inelastic reaction rates, we make the ansatz 
\begin{equation}
\Gamma(T)=\Gamma_0 \left( \frac{T}{T_0} \right)^n
\label{GammaT}
\end{equation}
and investigate different exponents, $n$=1-5. Following the width discussion at the end of the previous section, we employ the ranges 
$\Gmzero$=300-500\,MeV and $\Gtzero$=50-80\,MeV at $T_0$=180\,MeV for the molecular and tetraquark scenario, respectively.

\begin{figure}[!t]
\begin{minipage}[b]{1.0\linewidth}
\centering
\includegraphics[width=1.0\textwidth]{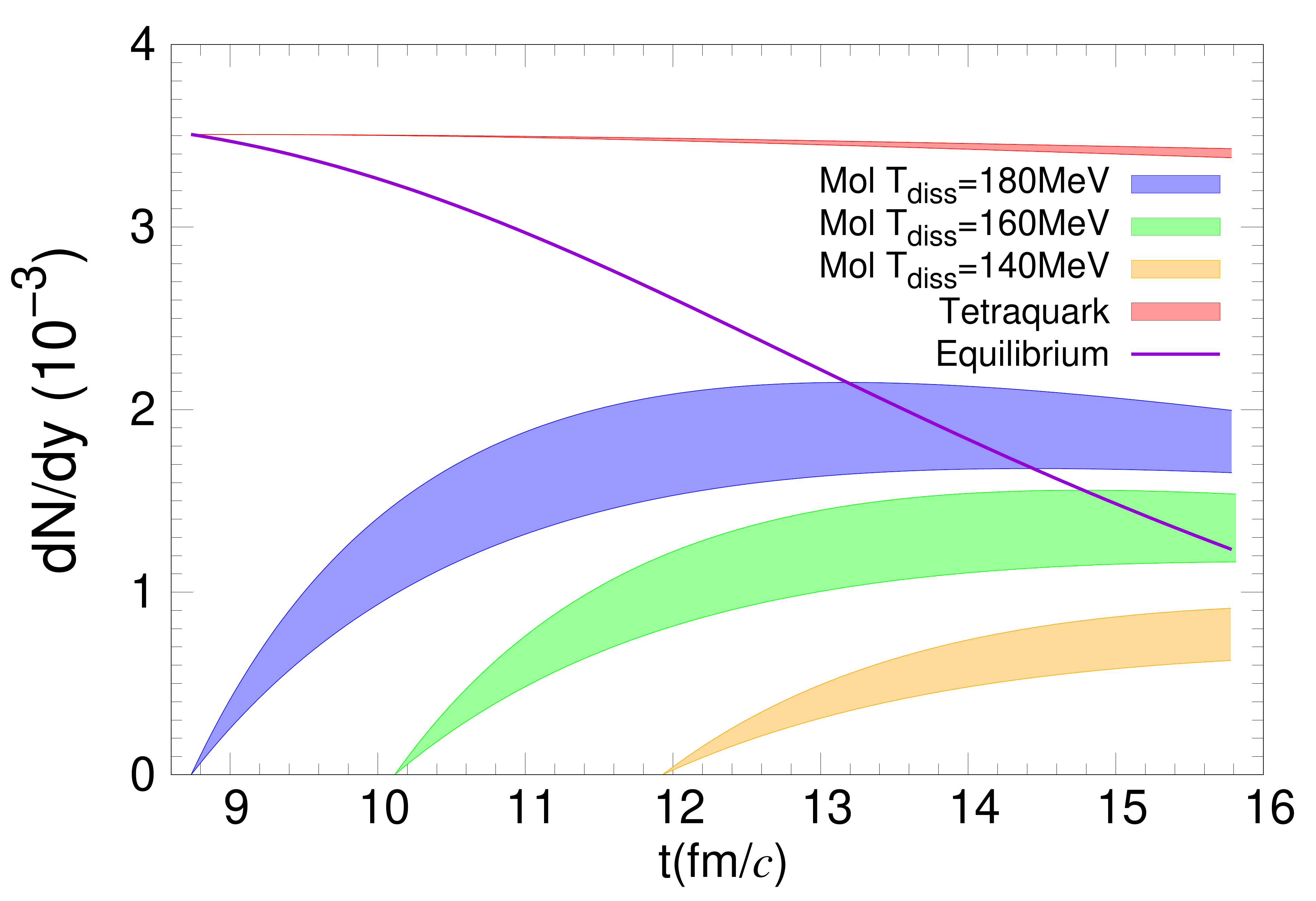}
\includegraphics[width=1.0\textwidth]{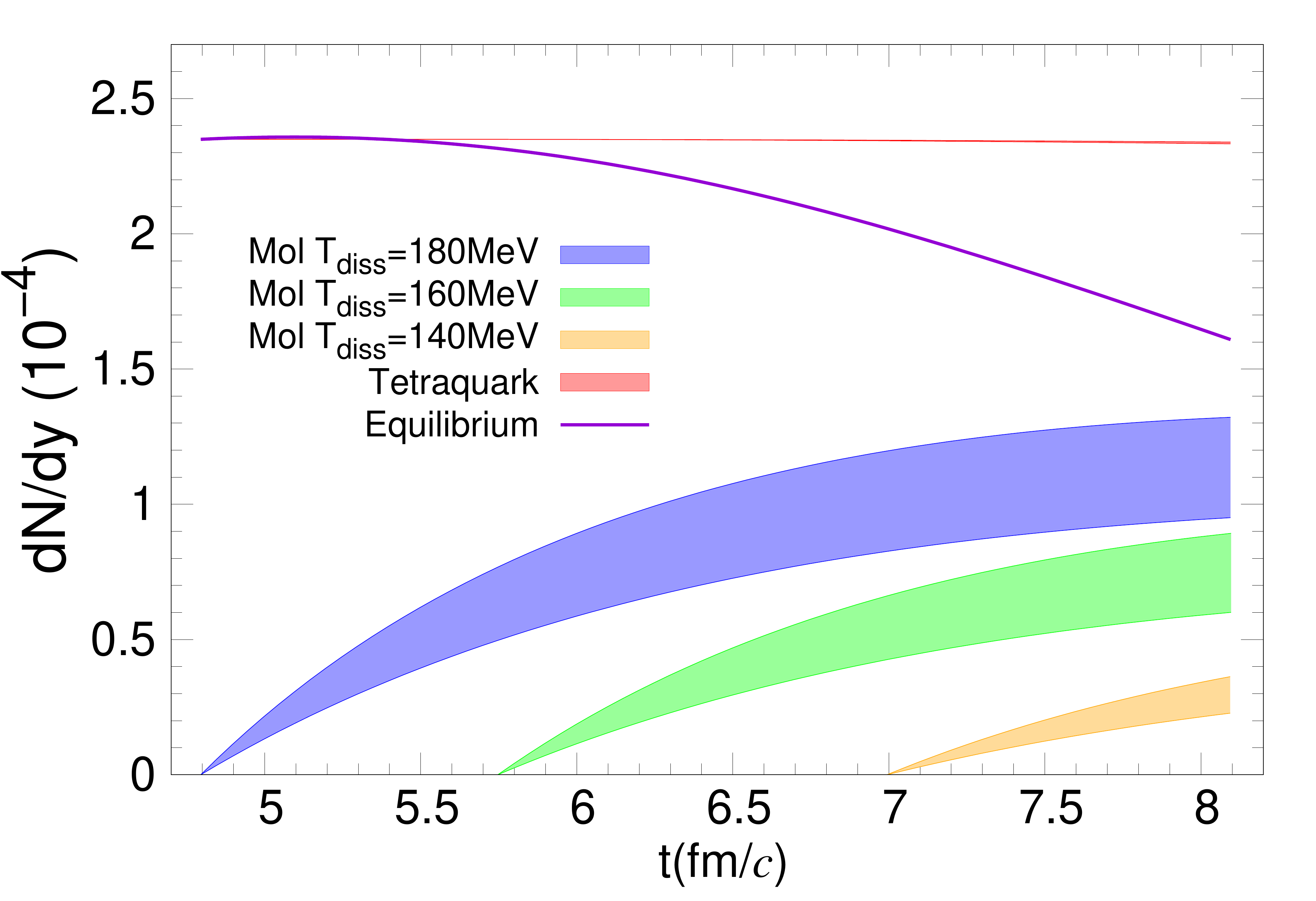}
\end{minipage}
\caption{Time evolution of the \X~equilibrium yields (solid line) and the solutions of the rate equation for the molecular (lower bands) and 
tetraquark (upper red band) scenarios for the hadronic phase in 0-20\% (upper panel) and 40-60\% (lower panel) Pb-Pb collisions at 5 TeV. 
The bands represent the width ranges with initial values of $\Gmzero$=300-500\,MeV and $\Gtzero$=50-80\,MeV with a temperature 
exponent of $n$=3; the blue, green and orange bands for the molecular scenario represent different onset temperatures for regeneration, 
corresponding to dissociation temperatures of $T_{\rm diss}$=180, 160 and 140\,MeV, respectively.} 
\label{fig_time-evo}
\end{figure}
We are now in position to solve the rate equation for \X~transport in the hadronic phase of the fireball evolution. The time evolution of 
the \X~yield per unit rapidity at mid-rapidity is plotted for the 2 scenarios, along with the temperature dependence of the equilibrium limit, 
in Fig.~\ref{fig_time-evo} for central and semi-central Pb-Pb collisions at 5.02\,TeV. Note that the equilibrium limit decreases with decreasing 
temperature; even though the charm-quark fugacity increases markedly toward lower temperatures, the thermal suppression of the \X~due 
its relatively large mass wins out (compared to the lower-mass open-charm states, like $D$ mesons, which mostly drive the value of 
$\gamma_c$). In the molecular scenario, with its large reaction rates, the transport evolution drives the \X~number rather close to equilibrium 
in the late stages of the evolution (more so in central collisions and for an earlier onset of the regeneration processes). On the other hand, 
even for our maximum estimate for the dissociation rate of the tetraquark state, its evolution in the hadronic phase is rather insignificant so 
that its yield stays close to the production level that it inherits from the QGP and its hadronization (which, as we argued above, should be 
reasonably close to the pertinent equilibrium limit in the transition region), similar to the results in Ref.~\cite{Cho:2013rpa}. As a consequence, 
the final yields of the tetraquark are about a factor of 2 larger than for the molecule, and even more so if the onset of regeneration 
for the molecule occurs at lower temperatures. 
This is qualitatively different from ICMs~\cite{Cho:2013rpa} where the larger size of the molecule configuration provides a much larger 
spatial phase space than for the tetraquark and thus generates appreciably larger yields. Even if we initialize the molecule at its equilibrium 
value at $T_0$, its suppression in the hadronic phase will still drive its yield (well) below the one of the tetraquark.

\begin{figure}[!t]
\begin{minipage}[b]{01.00\linewidth}
\centering
\includegraphics[width=1.0\textwidth]{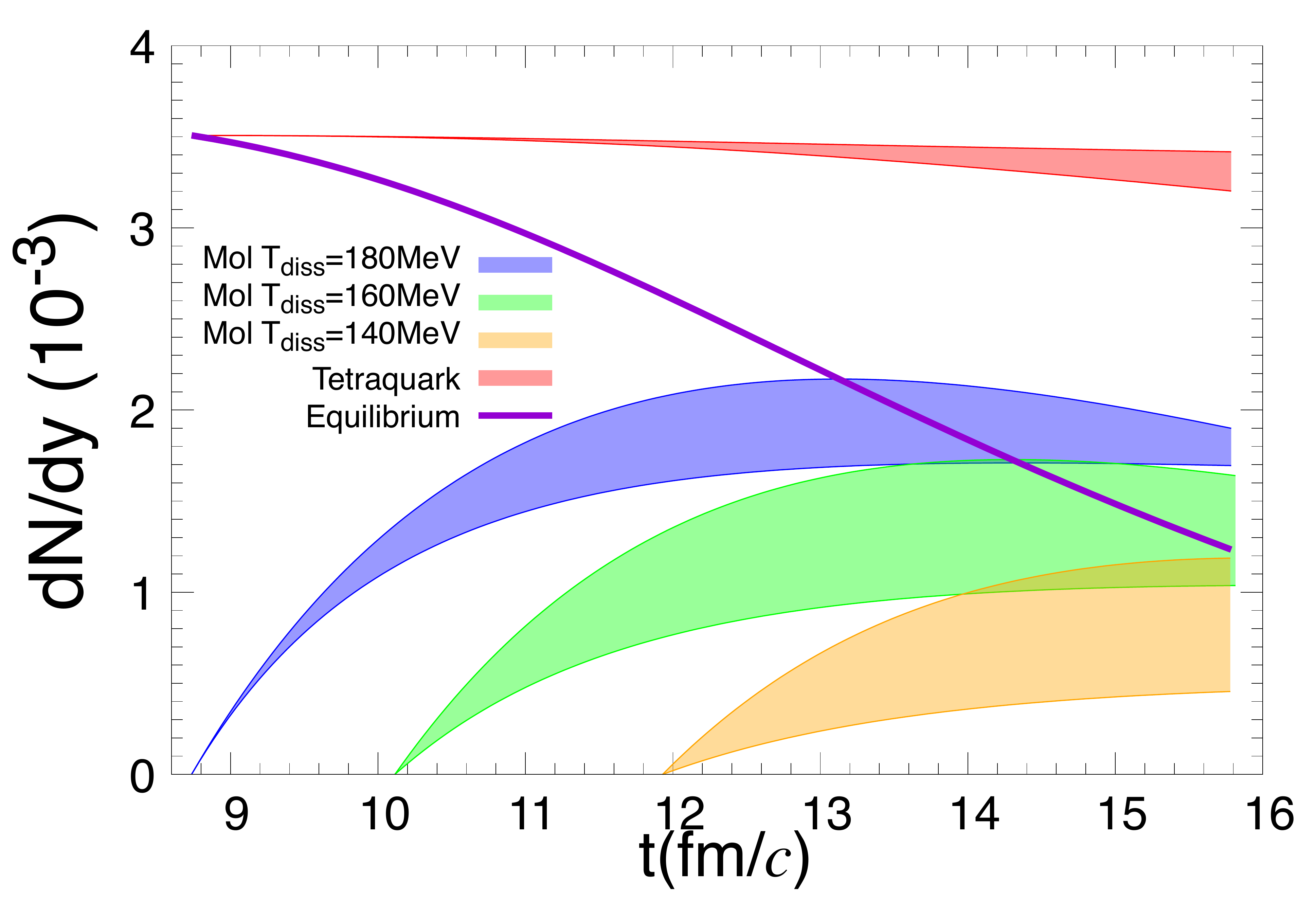}
\includegraphics[width=1.0\textwidth]{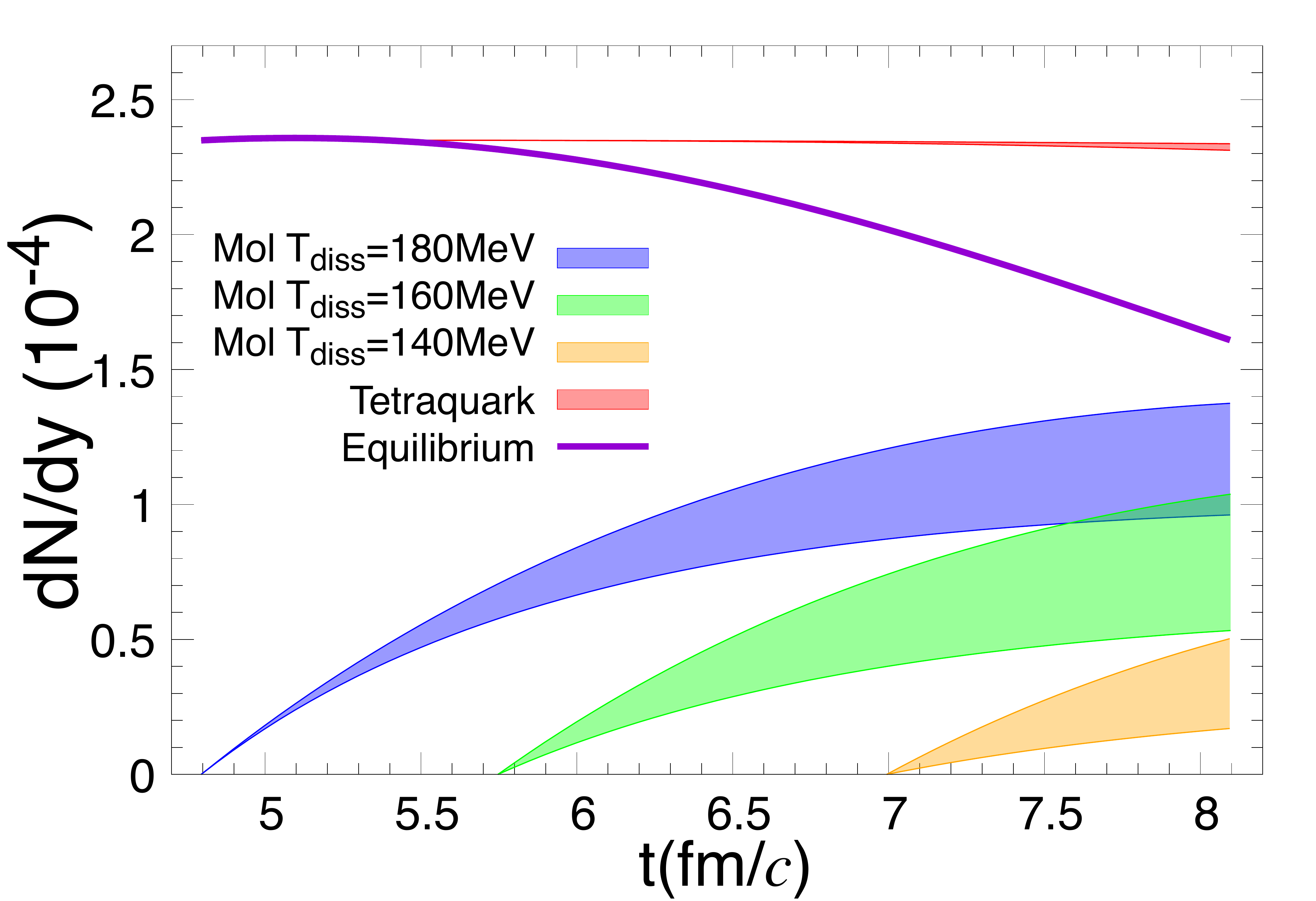}
\end{minipage}
\caption{Same as Fig.~\ref{fig_time-evo} but for fixed values of  $\Gtzero=65$\,MeV and $\Gmzero=400$\,MeV and varying the temperature 
   dependence of the widths according to Eq.~(\ref{GammaT}) with a range of $n$=1-5 for the exponent, resulting in the uncertainty bands for the
transport results.
}
\label{fig_n-exp}
\end{figure}

In Fig.~\ref{fig_n-exp} we illustrate the sensitivity of our calculations to the temperature dependence of the reaction rate by varying the 
exponent, $n$, in Eq.~(\ref{GammaT}). The resulting uncertainty 
systematics turn out to be rather similar to the variations of the maximal width values as displayed in Fig.~\ref{fig_time-evo}. Even for 
a rather large range of the exponent, $n$=1-5, the changes in the final \X~yields are quite moderate, 
\eg, $dN/dy = (1.15\pm 0.2)\cdot 10^{-4}$ for 40-60\% central collisions. In any case, the basic finding of a factor of $\sim$2 
smaller yield (or even more for smaller dissociation temperatures) for the molecular compared to the tetraquark scenario persists.

\section{\X~Observables}
\label{sec_obs}
We finally turn to the predictions of observables, specifically the centrality dependence of the \X~yields and their \pT spectra.
\begin{figure}[!t]
\begin{minipage}[b]{01.00\linewidth}
\centering
\includegraphics[width=1.0\textwidth]{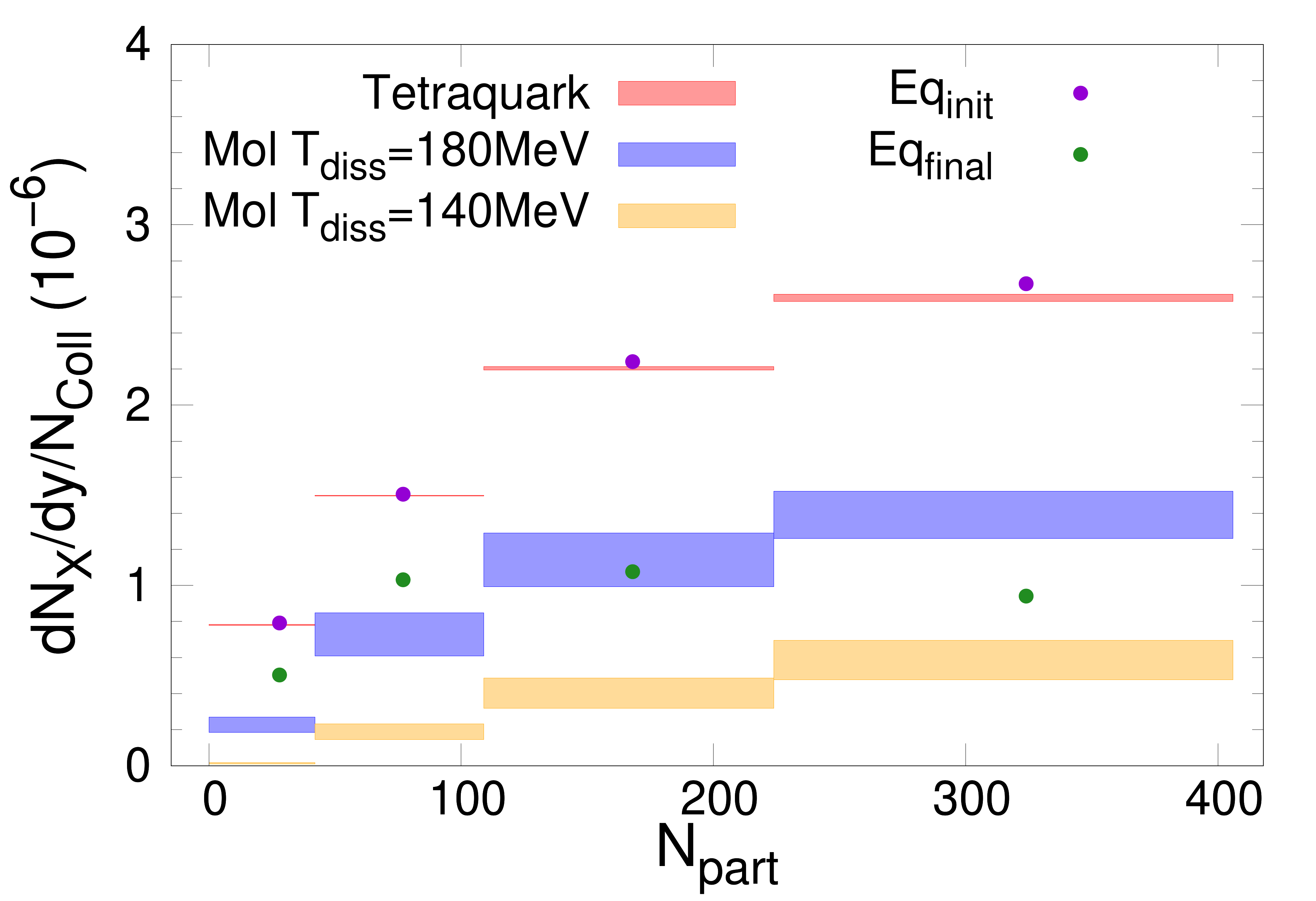}
\includegraphics[width=1.0\textwidth]{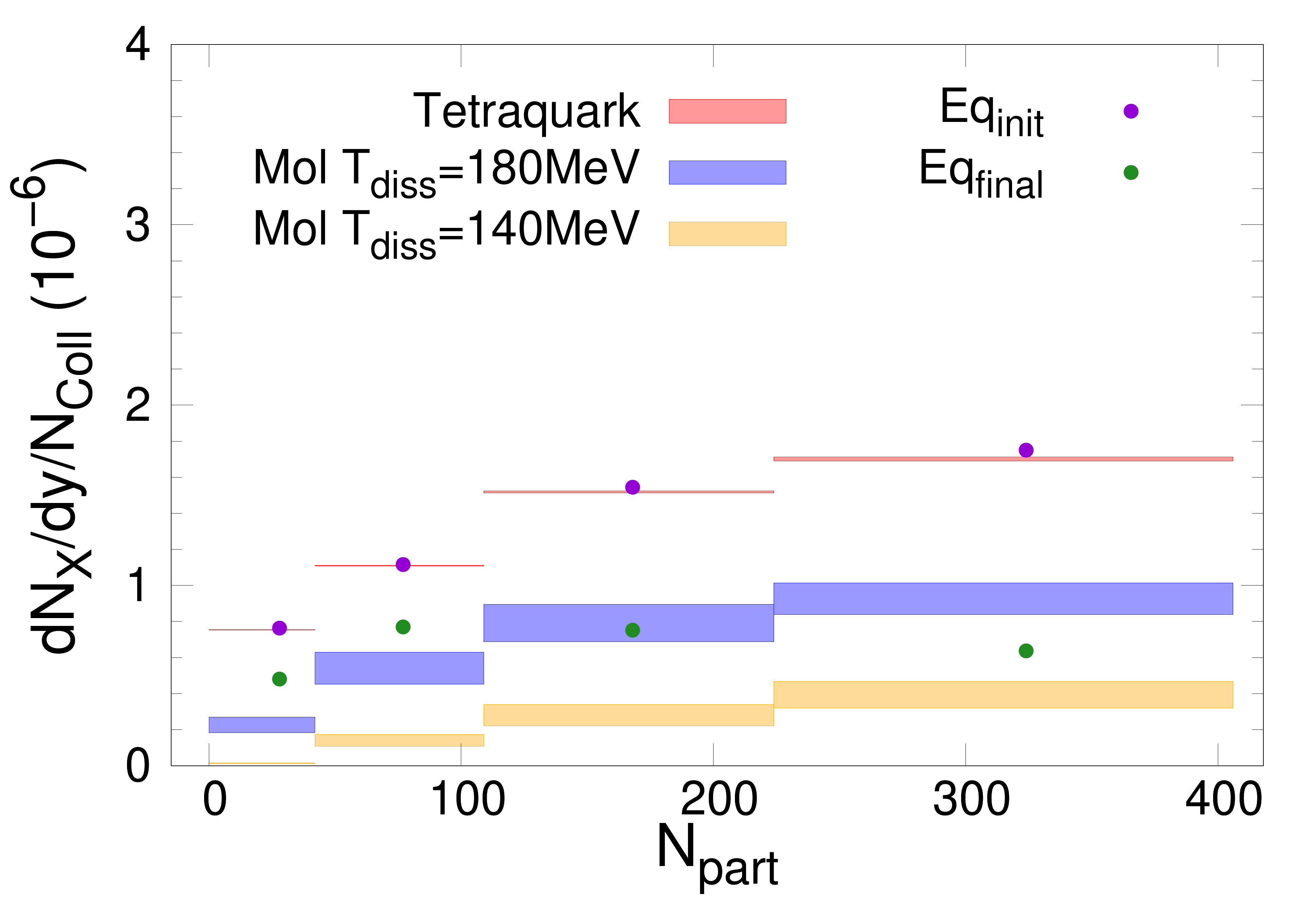}
\end{minipage}
\caption{Centrality dependence of the \X~production yields (vs.~number of participant nucleons, $N_{\rm part}$, in the collision), 
normalized to the number of primordial $NN$ collisions corresponding to each centrality class (red bars: tetraquark scenario, blue and orange bars: 
molecule scenario with \Td=180\,MeV and 140 MeV, respectively). Also shown are the values of the equilibrium limit at chemical (purple dots) 
and thermal freezeout (green dots). The results in the upper and lower panel are calculated without and with nuclear shadowing, 
respectively. The uncertainties in the transport results reflect our previously defined range of widths with $n$=3.}
\label{fig_centrality}
\end{figure}
The former are displayed for in Fig.~\ref{fig_centrality} for four centrality bins, and have been divided by the corresponding number of binary 
collisions, akin to a nuclear modification factor ($R_{\rm AA}$). The measurements of \X~production in $pp$ collisions are currently restricted
to the $J/\psi\pi\pi$ channel whose branching ratio is not well known, at approximately 5-30\%~\cite{Braaten:2019ags};
since we here calculate absolute production yields in AA collisions, we are not yet able to provide a meaningful $R_{\rm AA}$ result (although
the centrality dependence is unaffected by an overall normalization constant). We also note that our results do not include any ``primordial" 
production of \X~particles, but only the regeneration component. While the latter is expected to dominate the total yields, the former may play 
a role at high \pT, were thermal production drops off exponentially while the primordial one is expected of to be of power law type. 
With this in mind, we predict that the $R_{\rm AA}$ exhibits a significant rise with increasing centrality, with a production ratio of tetraquark 
over molecule scenarios of about 2, except for peripheral collisions (where it is larger), cf.~upper panel of Fig.~\ref{fig_centrality} .
When including an estimate of nuclear shadowing, where the total charm cross is assumed to drop by up to 20\% in central collisions, the total 
yields decrease by up to 40\%, driven by the $\gamma_c^2$ dependence of the equilibrium limit, rendering a somewhat less 
pronounced rise of the $R_{\rm AA}(N_{\rm part})$. 

Our results substantially differ from a recent calculation where a factor of $\sim$200 enhancement of \X~production in the molecular over the 
tetraquark scenario  was predicted for semi-/central Pb-Pb collisions at the LHC~\cite{Zhang:2020dwn}. The basic argument in that work is 
that the larger spatial size of the molecule (compared to the tetraquark) provides a much larger phase space for its production. However, it is 
not obvious how in such a picture the correct equilibrium limit is ensured, nor whether detailed balance is satisfied. These two principles are 
manifest in the kinetic rate equation (\ref{rate-eq}) used in the present work.

\begin{figure}[!t]
\begin{minipage}[b]{1.0\linewidth}
\centering
\includegraphics[width=1.0\textwidth]{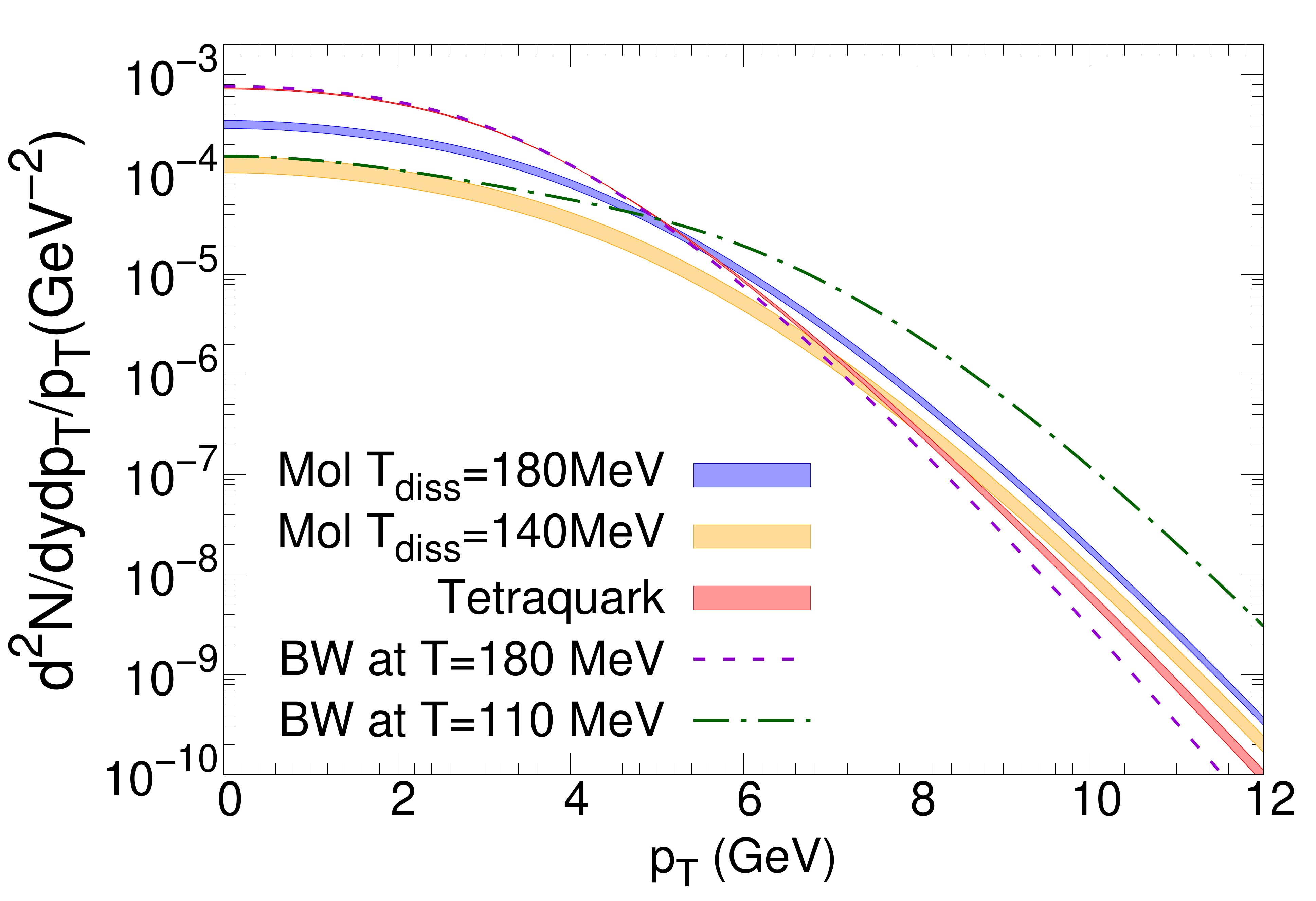}
\end{minipage}
\caption{Transverse-momentum spectra of the \X~in 0-20\% central Pb-Pb collisions for the molecular (blue band for \Td=180\,MeV 
and orange band for \Td=140\,MeV) and tetraquark (red band) scenarios, compared to blastwave spectra at chemical (dashed line) 
and thermal (dash-dotted line) freezeout. The width ranges and temperature exponent are as in Fig.~\ref{fig_centrality}, with no 
nuclear shadowing on the charm cross section included. }
\label{fig_pt}
\end{figure}
As an extra discriminator we investigate the \pT~spectra of the \X~in the different scenarios. Since the underlying open-charm hadron spectra 
in nuclear collisions at the LHC reach near thermal equilibrium at low \pT in central Pb-Pb collisions, we employ a thermal blastwave approximation 
resulting from the space-time profile of the fireball's expansion and temperature. Specifically, we weight the time evolution of the gain term of 
the rate equation with the time dependent blastwave expression for the \X and then renormalize the total \pT spectrum to the final yield 
obtained from the rate equation. The results are shown in Fig.~\ref{fig_pt}, together with the blastwave results for the equilibrium limit at 
chemical and thermal freezeout. As expected, the \pT~spectrum for the tetraquark is close to the blastwave at the hadronization temperature, 
while for the molecular scenario it is harder, although not by much if regeneration starts at $T_0$=180\,MeV, as most of the yield is still 
generated relatively early in the hadronic evolution. If the onset of regeneration is at lower temperatures, the hardening of the spectra is more 
pronounced.

\section{Conclusions}
\label{sec_concl}
We have investigated the production of the \X~particle in heavy-ion collisions using a thermal-rate equation approach, focusing on the 
hadronic phase of the fireball. We have found rather moderate differences in the yields within the two \X~structure scenarios, by around 
a factor of 2, which is smaller than in most coalescence model calculations which predict differences of up to two orders of magnitude. 
In our approach, the sensitivity to the internal structure is encoded in the reaction rate, which is expected to be much larger for the 
loosely bound hadronic molecule compared to the tetraquark as a compact bound state of colored anti-/diquarks. This implies that the 
yield of the molecule freezes out later in the hadronic evolution. Since the equilibrium limit decreases with temperature we expect a 
smaller yield for a molecule relative to a tetraquark, which is qualitatively different from coalescence models where the production phase 
space is largely driven by the spatial size of the \X~configuration. While the absolute yields of the \X~production depend quadratically on 
the charm-quark fugacity (which is an input to our approach that is beset with uncertainties due to the charm cross section in $pp$ collisions 
and nuclear shadowing), our findings for the ratio of the molecular over the tetraquark yield are independent of $\gamma_c$. 
We have also computed transverse-momentum spectra and found that they provide additional constraints on the production time in the 
fireball evolution, with harder spectra indicating later production. An open problem remains at which momenta the \X~production in 
heavy-ion collisions transits from kinetic production, as calculated in the present paper, to a $pp$-like power law shape characterizing 
the remnants of primordial emission (presumably from the fireball surface). This could be at a higher momentum scale than for light hadrons 
(even charmonia), due to the \X's fragile nature in the fireball while its thermal blastwave is rather susceptible to blue-shift effects in late-stage 
production (due to its large mass). If the \X~indeed turns out to be more of a tetraquark structure, it will be in order to scrutinize its transport 
in the sQGP near $T_c$ where anti-/diquark correlations are expected to emerge which then would have to fuse further into the \X~(and similarly 
exotic hadrons). In the present work we merely assumed this scenario to lead to a near-equilibrium yield at hadronization, which for peripheral 
collisions and at lower collision energies is likely to receive significant corrections.

\acknowledgments
This work has been supported by the U.S. National Science Foundation under grant no. PHY-1913286 and REU 
grant no. PHY-1659847, and by the TAMU Cyclotron Institute's Research Development (CIRD) program.

\bibliography{refcnew}

\begin{thebibliography}{29}
\expandafter\ifx\csname natexlab\endcsname\relax\def\natexlab#1{#1}\fi
\expandafter\ifx\csname bibnamefont\endcsname\relax
  \def\bibnamefont#1{#1}\fi
\expandafter\ifx\csname bibfnamefont\endcsname\relax
  \def\bibfnamefont#1{#1}\fi
\expandafter\ifx\csname citenamefont\endcsname\relax
  \def\citenamefont#1{#1}\fi
\expandafter\ifx\csname url\endcsname\relax
  \def\url#1{\texttt{#1}}\fi
\expandafter\ifx\csname urlprefix\endcsname\relax\def\urlprefix{URL }\fi
\providecommand{\bibinfo}[2]{#2}
\providecommand{\eprint}[2][]{\url{#2}}

\bibitem[{\citenamefont{Choi et~al.}(2003)}]{Choi:2003ue}
\bibinfo{author}{\bibfnamefont{S.}~\bibnamefont{Choi}} \bibnamefont{et~al.}
  (\bibinfo{collaboration}{Belle}), \bibinfo{journal}{Phys. Rev. Lett.}
  \textbf{\bibinfo{volume}{91}}, \bibinfo{pages}{262001}
  (\bibinfo{year}{2003}).

\bibitem[{\citenamefont{Esposito et~al.}(2017)\citenamefont{Esposito, Pilloni,
  and Polosa}}]{Esposito:2016noz}
\bibinfo{author}{\bibfnamefont{A.}~\bibnamefont{Esposito}},
  \bibinfo{author}{\bibfnamefont{A.}~\bibnamefont{Pilloni}}, \bibnamefont{and}
  \bibinfo{author}{\bibfnamefont{A.}~\bibnamefont{Polosa}},
  \bibinfo{journal}{Phys. Rept.} \textbf{\bibinfo{volume}{668}},
  \bibinfo{pages}{1} (\bibinfo{year}{2017}).

\bibitem[{\citenamefont{Hanhart et~al.}(2007)\citenamefont{Hanhart,
  Kalashnikova, Kudryavtsev, and Nefediev}}]{Hanhart:2007yq}
\bibinfo{author}{\bibfnamefont{C.}~\bibnamefont{Hanhart}},
  \bibinfo{author}{\bibfnamefont{Y.}~\bibnamefont{Kalashnikova}},
  \bibinfo{author}{\bibfnamefont{A.~E.} \bibnamefont{Kudryavtsev}},
  \bibnamefont{and} \bibinfo{author}{\bibfnamefont{A.}~\bibnamefont{Nefediev}},
  \bibinfo{journal}{Phys. Rev. D} \textbf{\bibinfo{volume}{76}},
  \bibinfo{pages}{034007} (\bibinfo{year}{2007}).

\bibitem[{\citenamefont{Molina and Oset}(2009)}]{Molina:2009ct}
\bibinfo{author}{\bibfnamefont{R.}~\bibnamefont{Molina}} \bibnamefont{and}
  \bibinfo{author}{\bibfnamefont{E.}~\bibnamefont{Oset}},
  \bibinfo{journal}{Phys. Rev. D} \textbf{\bibinfo{volume}{80}},
  \bibinfo{pages}{114013} (\bibinfo{year}{2009}).

\bibitem[{\citenamefont{Maiani et~al.}(2005)\citenamefont{Maiani, Piccinini,
  Polosa, and Riquer}}]{Maiani:2004vq}
\bibinfo{author}{\bibfnamefont{L.}~\bibnamefont{Maiani}},
  \bibinfo{author}{\bibfnamefont{F.}~\bibnamefont{Piccinini}},
  \bibinfo{author}{\bibfnamefont{A.}~\bibnamefont{Polosa}}, \bibnamefont{and}
  \bibinfo{author}{\bibfnamefont{V.}~\bibnamefont{Riquer}},
  \bibinfo{journal}{Phys. Rev. D} \textbf{\bibinfo{volume}{71}},
  \bibinfo{pages}{014028} (\bibinfo{year}{2005}).

\bibitem[{\citenamefont{Riek and Rapp}(2010)}]{Riek:2010fk}
\bibinfo{author}{\bibfnamefont{F.}~\bibnamefont{Riek}} \bibnamefont{and}
  \bibinfo{author}{\bibfnamefont{R.}~\bibnamefont{Rapp}},
  \bibinfo{journal}{Phys. Rev. C} \textbf{\bibinfo{volume}{82}},
  \bibinfo{pages}{035201} (\bibinfo{year}{2010}).

\bibitem[{\citenamefont{{CMS Collaboration}}(2019)}]{CMS:2019vma}
\bibinfo{author}{\bibnamefont{{CMS Collaboration}}}
  (\bibinfo{collaboration}{CMS}), \bibinfo{type}{CMS Physics Analysis Summary}
  \bibinfo{number}{CMS-PAS-HIN-19-005} (\bibinfo{year}{2019}).

\bibitem[{\citenamefont{Cho et~al.}(2011)}]{Cho:2010db}
\bibinfo{author}{\bibfnamefont{S.}~\bibnamefont{Cho}} \bibnamefont{et~al.}
  (\bibinfo{collaboration}{ExHIC}), \bibinfo{journal}{Phys. Rev. Lett.}
  \textbf{\bibinfo{volume}{106}}, \bibinfo{pages}{212001}
  (\bibinfo{year}{2011}).

\bibitem[{\citenamefont{Sun and Chen}(2017)}]{Sun:2017ooe}
\bibinfo{author}{\bibfnamefont{K.-J.} \bibnamefont{Sun}} \bibnamefont{and}
  \bibinfo{author}{\bibfnamefont{L.-W.} \bibnamefont{Chen}},
  \bibinfo{journal}{Phys. Rev. C} \textbf{\bibinfo{volume}{95}},
  \bibinfo{pages}{044905} (\bibinfo{year}{2017}).

\bibitem[{\citenamefont{Fontoura et~al.}(2019)\citenamefont{Fontoura, Krein,
  Valcarce, and Vijande}}]{Fontoura:2019opw}
\bibinfo{author}{\bibfnamefont{C.}~\bibnamefont{Fontoura}},
  \bibinfo{author}{\bibfnamefont{G.}~\bibnamefont{Krein}},
  \bibinfo{author}{\bibfnamefont{A.}~\bibnamefont{Valcarce}}, \bibnamefont{and}
  \bibinfo{author}{\bibfnamefont{J.}~\bibnamefont{Vijande}},
  \bibinfo{journal}{Phys. Rev. D} \textbf{\bibinfo{volume}{99}},
  \bibinfo{pages}{094037} (\bibinfo{year}{2019}).

\bibitem[{\citenamefont{Zhang et~al.}(2021)\citenamefont{Zhang, Liao, Wang,
  Wang, and Xing}}]{Zhang:2020dwn}
\bibinfo{author}{\bibfnamefont{H.}~\bibnamefont{Zhang}},
  \bibinfo{author}{\bibfnamefont{J.}~\bibnamefont{Liao}},
  \bibinfo{author}{\bibfnamefont{E.}~\bibnamefont{Wang}},
  \bibinfo{author}{\bibfnamefont{Q.}~\bibnamefont{Wang}}, \bibnamefont{and}
  \bibinfo{author}{\bibfnamefont{H.}~\bibnamefont{Xing}},
  \bibinfo{journal}{Phys. Rev. Lett.} \textbf{\bibinfo{volume}{126}},
  \bibinfo{pages}{012301} (\bibinfo{year}{2021}).

\bibitem[{\citenamefont{Ravagli and Rapp}(2007)}]{Ravagli:2007xx}
\bibinfo{author}{\bibfnamefont{L.}~\bibnamefont{Ravagli}} \bibnamefont{and}
  \bibinfo{author}{\bibfnamefont{R.}~\bibnamefont{Rapp}},
  \bibinfo{journal}{Phys. Lett. B} \textbf{\bibinfo{volume}{655}},
  \bibinfo{pages}{126} (\bibinfo{year}{2007}).

\bibitem[{\citenamefont{Andronic et~al.}(2019)\citenamefont{Andronic,
  Braun-Munzinger, K{\"o}hler, Redlich, and Stachel}}]{Andronic:2019wva}
\bibinfo{author}{\bibfnamefont{A.}~\bibnamefont{Andronic}},
  \bibinfo{author}{\bibfnamefont{P.}~\bibnamefont{Braun-Munzinger}},
  \bibinfo{author}{\bibfnamefont{M.~K.} \bibnamefont{K{\"o}hler}},
  \bibinfo{author}{\bibfnamefont{K.}~\bibnamefont{Redlich}}, \bibnamefont{and}
  \bibinfo{author}{\bibfnamefont{J.}~\bibnamefont{Stachel}},
  \bibinfo{journal}{Phys. Lett. B} \textbf{\bibinfo{volume}{797}},
  \bibinfo{pages}{134836} (\bibinfo{year}{2019}).

\bibitem[{\citenamefont{Grandchamp et~al.}(2004)\citenamefont{Grandchamp, Rapp,
  and Brown}}]{Grandchamp:2003uw}
\bibinfo{author}{\bibfnamefont{L.}~\bibnamefont{Grandchamp}},
  \bibinfo{author}{\bibfnamefont{R.}~\bibnamefont{Rapp}}, \bibnamefont{and}
  \bibinfo{author}{\bibfnamefont{G.~E.} \bibnamefont{Brown}},
  \bibinfo{journal}{Phys. Rev. Lett.} \textbf{\bibinfo{volume}{92}},
  \bibinfo{pages}{212301} (\bibinfo{year}{2004}).

\bibitem[{\citenamefont{Zhao and Rapp}(2011)}]{Zhao:2011cv}
\bibinfo{author}{\bibfnamefont{X.}~\bibnamefont{Zhao}} \bibnamefont{and}
  \bibinfo{author}{\bibfnamefont{R.}~\bibnamefont{Rapp}},
  \bibinfo{journal}{Nucl. Phys. A} \textbf{\bibinfo{volume}{859}},
  \bibinfo{pages}{114} (\bibinfo{year}{2011}).

\bibitem[{\citenamefont{Du et~al.}(2017)\citenamefont{Du, Rapp, and
  He}}]{Du:2017qkv}
\bibinfo{author}{\bibfnamefont{X.}~\bibnamefont{Du}},
  \bibinfo{author}{\bibfnamefont{R.}~\bibnamefont{Rapp}}, \bibnamefont{and}
  \bibinfo{author}{\bibfnamefont{M.}~\bibnamefont{He}}, \bibinfo{journal}{Phys.
  Rev. C} \textbf{\bibinfo{volume}{96}}, \bibinfo{pages}{054901}
  (\bibinfo{year}{2017}).

\bibitem[{\citenamefont{Cho and Lee}(2013)}]{Cho:2013rpa}
\bibinfo{author}{\bibfnamefont{S.}~\bibnamefont{Cho}} \bibnamefont{and}
  \bibinfo{author}{\bibfnamefont{S.~H.} \bibnamefont{Lee}},
  \bibinfo{journal}{Phys. Rev. C} \textbf{\bibinfo{volume}{88}},
  \bibinfo{pages}{054901} (\bibinfo{year}{2013}).

\bibitem[{\citenamefont{Abreu et~al.}(2016)\citenamefont{Abreu, Khemchandani,
  Martinez~Torres, Navarra, and Nielsen}}]{Abreu:2016qci}
\bibinfo{author}{\bibfnamefont{L.}~\bibnamefont{Abreu}},
  \bibinfo{author}{\bibfnamefont{K.}~\bibnamefont{Khemchandani}},
  \bibinfo{author}{\bibfnamefont{A.}~\bibnamefont{Martinez~Torres}},
  \bibinfo{author}{\bibfnamefont{F.}~\bibnamefont{Navarra}}, \bibnamefont{and}
  \bibinfo{author}{\bibfnamefont{M.}~\bibnamefont{Nielsen}},
  \bibinfo{journal}{Phys. Lett. B} \textbf{\bibinfo{volume}{761}},
  \bibinfo{pages}{303} (\bibinfo{year}{2016}).

\bibitem[{\citenamefont{Hong et~al.}(2018)\citenamefont{Hong, Cho, Song, and
  Lee}}]{Hong:2018mpk}
\bibinfo{author}{\bibfnamefont{J.}~\bibnamefont{Hong}},
  \bibinfo{author}{\bibfnamefont{S.}~\bibnamefont{Cho}},
  \bibinfo{author}{\bibfnamefont{T.}~\bibnamefont{Song}}, \bibnamefont{and}
  \bibinfo{author}{\bibfnamefont{S.~H.} \bibnamefont{Lee}},
  \bibinfo{journal}{Phys. Rev. C} \textbf{\bibinfo{volume}{98}},
  \bibinfo{pages}{014913} (\bibinfo{year}{2018}).

\bibitem[{\citenamefont{Du and Rapp}(2015)}]{Du:2015wha}
\bibinfo{author}{\bibfnamefont{X.}~\bibnamefont{Du}} \bibnamefont{and}
  \bibinfo{author}{\bibfnamefont{R.}~\bibnamefont{Rapp}},
  \bibinfo{journal}{Nucl. Phys. A} \textbf{\bibinfo{volume}{943}},
  \bibinfo{pages}{147} (\bibinfo{year}{2015}).

\bibitem[{\citenamefont{Acharya et~al.}(2018)}]{Acharya:2017kfy}
\bibinfo{author}{\bibfnamefont{S.}~\bibnamefont{Acharya}} \bibnamefont{et~al.}
  (\bibinfo{collaboration}{ALICE}), \bibinfo{journal}{JHEP}
  \textbf{\bibinfo{volume}{04}}, \bibinfo{pages}{108} (\bibinfo{year}{2018}).

\bibitem[{\citenamefont{He and Rapp}(2019)}]{He:2019tik}
\bibinfo{author}{\bibfnamefont{M.}~\bibnamefont{He}} \bibnamefont{and}
  \bibinfo{author}{\bibfnamefont{R.}~\bibnamefont{Rapp}},
  \bibinfo{journal}{Phys. Lett. B} \textbf{\bibinfo{volume}{795}},
  \bibinfo{pages}{117} (\bibinfo{year}{2019}).

\bibitem[{\citenamefont{Cleven et~al.}(2019)\citenamefont{Cleven, Magas, and
  Ramos}}]{Cleven:2019cre}
\bibinfo{author}{\bibfnamefont{M.}~\bibnamefont{Cleven}},
  \bibinfo{author}{\bibfnamefont{V.~K.} \bibnamefont{Magas}}, \bibnamefont{and}
  \bibinfo{author}{\bibfnamefont{A.}~\bibnamefont{Ramos}},
  \bibinfo{journal}{Phys. Lett. B} \textbf{\bibinfo{volume}{799}},
  \bibinfo{pages}{135050} (\bibinfo{year}{2019}).

\bibitem[{\citenamefont{Fuchs et~al.}(2006)\citenamefont{Fuchs, Martemyanov,
  Faessler, and Krivoruchenko}}]{Fuchs:2004fh}
\bibinfo{author}{\bibfnamefont{C.}~\bibnamefont{Fuchs}},
  \bibinfo{author}{\bibfnamefont{B.}~\bibnamefont{Martemyanov}},
  \bibinfo{author}{\bibfnamefont{A.}~\bibnamefont{Faessler}}, \bibnamefont{and}
  \bibinfo{author}{\bibfnamefont{M.}~\bibnamefont{Krivoruchenko}},
  \bibinfo{journal}{Phys. Rev. C} \textbf{\bibinfo{volume}{73}},
  \bibinfo{pages}{035204} (\bibinfo{year}{2006}).

\bibitem[{\citenamefont{He et~al.}(2011)\citenamefont{He, Fries, and
  Rapp}}]{He:2011yi}
\bibinfo{author}{\bibfnamefont{M.}~\bibnamefont{He}},
  \bibinfo{author}{\bibfnamefont{R.~J.} \bibnamefont{Fries}}, \bibnamefont{and}
  \bibinfo{author}{\bibfnamefont{R.}~\bibnamefont{Rapp}},
  \bibinfo{journal}{Phys. Lett. B} \textbf{\bibinfo{volume}{701}},
  \bibinfo{pages}{445} (\bibinfo{year}{2011}).

\bibitem[{\citenamefont{Brazzi et~al.}(2011)\citenamefont{Brazzi, Grinstein,
  Piccinini, Polosa, and Sabelli}}]{Brazzi:2011fq}
\bibinfo{author}{\bibfnamefont{F.}~\bibnamefont{Brazzi}},
  \bibinfo{author}{\bibfnamefont{B.}~\bibnamefont{Grinstein}},
  \bibinfo{author}{\bibfnamefont{F.}~\bibnamefont{Piccinini}},
  \bibinfo{author}{\bibfnamefont{A.~D.} \bibnamefont{Polosa}},
  \bibnamefont{and} \bibinfo{author}{\bibfnamefont{C.}~\bibnamefont{Sabelli}},
  \bibinfo{journal}{Phys. Rev. D} \textbf{\bibinfo{volume}{84}},
  \bibinfo{pages}{014003} (\bibinfo{year}{2011}).

\bibitem[{\citenamefont{Ferreiro and Lansberg}(2018)}]{Ferreiro:2018wbd}
\bibinfo{author}{\bibfnamefont{E.~G.} \bibnamefont{Ferreiro}} \bibnamefont{and}
  \bibinfo{author}{\bibfnamefont{J.-P.} \bibnamefont{Lansberg}},
  \bibinfo{journal}{JHEP} \textbf{\bibinfo{volume}{10}}, \bibinfo{pages}{094}
  (\bibinfo{year}{2018}), \bibinfo{note}{[Erratum: JHEP 03, 063 (2019)]}.

\bibitem[{\citenamefont{Shuryak and Zahed}(2004)}]{Shuryak:2003ja}
\bibinfo{author}{\bibfnamefont{E.}~\bibnamefont{Shuryak}} \bibnamefont{and}
  \bibinfo{author}{\bibfnamefont{I.}~\bibnamefont{Zahed}},
  \bibinfo{journal}{Phys. Rev. D} \textbf{\bibinfo{volume}{69}},
  \bibinfo{pages}{046005} (\bibinfo{year}{2004}).

\bibitem[{\citenamefont{Braaten et~al.}(2019)\citenamefont{Braaten, He, and
  Ingles}}]{Braaten:2019ags}
\bibinfo{author}{\bibfnamefont{E.}~\bibnamefont{Braaten}},
  \bibinfo{author}{\bibfnamefont{L.-P.} \bibnamefont{He}}, \bibnamefont{and}
  \bibinfo{author}{\bibfnamefont{K.}~\bibnamefont{Ingles}}
  (\bibinfo{year}{2019}), \eprint{1908.02807}.

\end{thebibliography}

\end{document}